\documentclass[12pt]{article}

\usepackage{amsmath,amssymb,amsfonts}

\begin{document}

\title{{\Large{\bf A variant of Wiener's attack on RSA}}}

\author{Andrej Dujella}

\date{}
\maketitle

 \vspace{1ex}

\begin{abstract}
Wiener's attack is a well-known polynomial-time attack on a RSA
cryptosystem with small secret decryption exponent $d$, which works
if $d<n^{0.25}$, where $n=pq$ is the modulus of the cryptosystem.
Namely, in that case, $d$ is the denominator of some convergent
$p_m/q_m$ of the continued fraction expansion of $e/n$, and therefore $d$
can be computed efficiently from the public key $(n,e)$.

There are several extensions of Wiener's attack that allow the RSA
cryptosystem to be broken when $d$ is a few bits longer than $n^{0.25}$.
They all have the run-time complexity (at least) $O(D^2)$, where
$d=Dn^{0.25}$. Here we propose a new variant of Wiener's attack, which
uses results on Diophantine approximations of the form $|\alpha - p/q|
< c/q^2$, and ``meet-in-the-middle'' variant for testing the candidates
(of the form $rq_{m+1} + sq_m$) for the secret exponent. This
decreases the run-time complexity of the attack to $O(D\log(D))$ (with
the space complexity $O(D)$).
\end{abstract}

\section{Introduction}

The most popular public key cryptosystem in use today is
the RSA cryptosystem, introduced by Rivest, Shamir, and Adleman \cite{RSA}.
Its security is based on the intractability of
the integer factorization problem.

The modulus $n$ of a RSA cryptosystem is the product of two large
primes $p$ and $q$. The public exponent $e$ and the secret
exponent $d$ are related by
\begin{equation} \label{phi1}
ed\equiv 1 \pmod{\varphi(n)},
\end{equation}
where $\varphi(n)= (p-1)(q-1)$. In a typical RSA cryptosystem,
$p$ and $q$ have approximately the same number of bits, while $e<n$.
The encryption and decryption algorithms are given by $C= M^e
\bmod n$, $M=C^d \bmod n$.

To speed up the RSA decryption one may try to use small secret
decryption exponent $d$. The choice of a small $d$ is especially
interesting when there is a large difference in computing power
between two communicating devices, e.g. in communication between
a smart card and a larger computer.
In this situation, it would be desirable that the smart card has a small secret exponent,
while the larger computer has a small public exponent, to reduce the processing
required in the smart card.

In 1990, Wiener \cite{Wiener} described a polynomial time
algorithm for breaking a typical (i.e. $p$ and $q$ are of the
same size and $e<n$) RSA cryptosystem if the secret exponent $d$ has at most
one-quarter as many bits as the modulus $n$.
From (\ref{phi1}) it follows that there is an
integer $k$ such that $ed- k\varphi(n)=1$. Since $ \varphi(n)
\approx n$, we have that $\frac{k}{d} \approx \frac{e}{n}$.
Wiener's attack is usually described in the following form (see
\cite{B-notices,Smart}):

\emph{If $p<q<2p$, $e<n$ and $d<\frac{1}{3}\sqrt[4]{n}$, then $d$ is the
denominator of some convergent of the continued fraction expansion of
$\frac{e}{n}$.}

Indeed, under these assumptions it is easy to show that
$$ \left|\frac{e}{n} - \frac{k}{d} \right| < \frac{1}{2d^2}. $$
By the classical Legendre's theorem, $\frac{k}{d}$ is
some convergent $\frac{p_m}{q_m}$ of the continued fraction expansion of $\frac{e}{n}$, and
therefore $d$ can be computed efficiently from the public key
$(n,e)$. Namely, the total number of convergents is of order $O(\log n)$, and each convergent can be
tested in polynomial time.

In 1997,  Verheul and van Tilborg \cite{V-vT} proposed an extension
of Wiener's attack that allows the RSA cryptosystem to be broken when $d$ is a
few bits longer than $n^{0.25}$. For $d>n^{0.25}$ their attack
needs to do an exhaustive search for about $2t+8$ bits (under
reasonable assumptions on involved partial convergents), where
$t=\log_2(d/n^{0.25})$.

\medskip

In \cite{DujeRSA}, we proposed a slight
modification of the Verheul and van Tilborg attack, based on
Worley's result on Diophantine approximations \cite{Worley},
which implies that all rationals $\frac{p}{q}$
satisfying the inequality
\begin{equation} \label{eq:zzz}
 \left| \alpha - \frac{p}{q} \right| < \frac{c}{q^2},
\end{equation}
for a positive real number $c$,
have the form
\begin{equation}  \label{eq:zzzz}
\frac{p}{q} = \frac{rp_{m+1} \pm sp_m}{ rq_{m+1} \pm sq_m }
\end{equation}
 for some $m\geq -1$
and nonnegative integers $r$ and $s$ such that $rs < 2c$.
It has been shown recently in \cite{DuIbr} that Worley's result is sharp,
in the sense that the condition $rs<2c$ cannot be replaced by
$rs < (2-\varepsilon)c$ for any $\varepsilon$.

In both mentioned extensions of Wiener's attack, the candidates
for the secret exponent are of the form $d = r q_{m+1} + s q_m$.
Then we test all possibilities for $d$. The number of possibilities is
roughly the product of the number of possibilities for $r$ and the number of
possibilities for $s$, which is $O(D^2)$, where $d=D n^{0.25}$.
More precisely, the number of possible pairs $(r,s)$ in the Verheul and van Tilborg attack
is $O(D^2A^2)$, where $A=\max\{a_i : i=m\!+\!1,m\!+\!2,m\!+\!3\}$, while in our
variant the number of pairs is $O(D^2\log A)$ (and also $O(D^2\log D)$).

Another modification
of the Verheul and van Tilborg attack has been recently proposed by Sun, Wu an Chen \cite{SWC}.
It requires (heuristically) an exhaustive search for about $2t-10$ bits, so its complexity
is also $O(D^2)$.
We cannot expect drastic improvements here, since, by a result of
Steinfeld, Contini, Wang and Pieprzyk \cite{SCWP},
there does not exist an attack in this class with subexponential running time.

\medskip

Boneh and Durfee \cite{B-D1} and Bl\"omer and May \cite{B-M}
proposed attacks based on Coppersmith's lattice-based technique for
finding small roots of modular polynomials equations using LLL-algorithm. The attacks work
if $d< n^{0.292}$.
The conjecture is that the right bound below which a typical version of RSA
is insecure is $d< n^{0.5}$.

\medskip

In the present paper, we propose a new variant of Wiener's attack. It also uses
continued fractions and searches for candidates for the secret key in the form
$d=rq_{m+1}+sq_{m}$. However, the searching phase of this variant is significantly faster.
Its complexity is $O(D\log{D})$,
and it works efficiently for $d<10^{30}n^{0.25}$.
Although this bound is asymptotically weaker than the bounds in the above mentioned
attacks based on the LLL-algorithm (note however that these bounds are not strictly proved since
Coppersmith's theorem in the bivariate case is only a heuristic result -
see also \cite{Hinek,HLT}), for practical values of $n$ (e.g. for 1024-bits) these bounds
are of comparable size.

\section{The Verheul and van Tilborg attack}

In this section we briefly describe
the Verheul and van Tilborg attack \cite{V-vT} and its modification from \cite{DujeRSA}.

We assume that $p<q<2p$ and $e<n$.
Then it is easy to see that
\begin{equation} \label{eq:nej1}
\left| \frac{e}{n} - \frac{k}{d} \right| < \frac{2.122\, e}{n\sqrt{n}}.
\end{equation}
Let $m$ be the largest (odd) integer satisfying
$\frac{p_m}{q_m} - \frac{e}{n} > \frac{2.122\, e}{n\sqrt{n}}$.
Verheul and van Tilborg proposed to search for $\frac{k}{d}$ among the fractions of the form
$\frac{r p_{m+1}+s p_m}{rq_{m+1}+ s q_m}$. This leads to the system
\begin{eqnarray*}
r p_{m+1}+s p_m &=& k, \\
r q_{m+1}+ s q_m &=& d.
\end{eqnarray*}
The determinant of the system satisfies
$|p_{m+1}q_m - q_{m+1}p_m| = 1$, and therefore
the system has (positive) integer solutions:
\begin{eqnarray*}
r &=& dp_{m} -k q_m,  \\
s &=& kq_{m+1}- d p_{m+1} .
\end{eqnarray*}
If $r$ and $s$ are small, then they can be found
by an exhaustive search. Let $[a_0;a_1,a_2,\ldots]$ be the continued
fraction expansion of $e/n$ and $D=d/n^{0.25}$.
In \cite{DujeRSA}, the following upper bounds for $r$ and $s$ were derived:
\begin{eqnarray*}
r &<& \max \{ \sqrt{2.122(a_{m+3}+2)}(a_{m+2}+1)D,\, \sqrt{2.122(a_{m+2}+2)}D \}, \\
s &<& \max \{ 2\sqrt{2.122(a_{m+3}+2)}D,\,\sqrt{2.122(a_{m+2}+2)}(a_{m+1}+1)D \} .
\end{eqnarray*}

The modified attack proposed in \cite{DujeRSA} searches
for $\frac{k}{d}$ among the fractions of the forms
$\frac{r p_{m+1}+s p_m}{rq_{m+1}+ s q_m}$,
$\frac{r p_{m+2}-s p_{m+1}}{rq_{m+2}- s q_{m+1}}$ and
$\frac{r p_{m+3}+s p_{m+2}}{rq_{m+3}+ s q_{m+2}}$.
It results with bounds for $r$ and $s$ which are (almost)
independent on the partial quotients $a_m$'s.
Hence, in both attacks bounds for $r$ and $s$ are of the form
$O(D)$, but in the case of \cite{DujeRSA} the implied constants
are much smaller (indeed, the table in Section \ref{impl} shows
that with high probability we have $r<4D$ and $s<4D$).

\section{Testing the candidates}

 There are two principal methods for testing candidates for the secret exponent $d$.

{\bf Method I} (\cite{Wiener}): Compute $p$ and $q$, assuming $d$ is the correct guess,
using the following formulas:
$$ \varphi(n)=(de-1)/k, \quad p+q=n+1-\varphi(n), $$
$$ (q-p)^2 = (p+q)^2 - 4n, $$
$$ p = \frac{p+q}{2} - \frac{q-p}{2}, \quad q = \frac{p+q}{2} + \frac{q-p}{2}. $$

{\bf Method II}  (\cite[Chapter 17]{Smart}): Test the congruence $(M^e)^d \equiv M \!\!\pmod{n}$,
for some random value of $M$, or simply for $M=2$.

\medskip

Both methods are very efficient. But in the situation where we have to test huge amount of
candidates for $d$ of the form $rq_{m+1}+sq_{m}$, there is a significant difference between them.
With the Method I it seems that we cannot avoid testing separately all possible pairs $(r,s)$.
On the other hand, here we present a new idea, which is to apply
``meet-in-the-middle'' to the Method II.

We want to test whether
\begin{equation} \label{eq:z}
 2^{e(rq_{m+1}+sq_m)} \equiv 2 \pmod{n}.
\end{equation}
Note that $m$ is (almost) fixed. Indeed, let $m'$ be the largest odd integer such that
$$ \frac{p_{m'}}{q_{m'}} > \frac{e}{n} + \frac{2.122 e}{n\sqrt{n}}. $$
Then $m\in \{m', m'+1, m'+2 \}$ (see \cite{DujeRSA} for details).

Let $2^{e q_{m+1}}
\bmod{n} = a$, $(2^{e q_m})^{-1} \bmod{n} = b$. Then we test the
congruence
\begin{equation} \label{eq:zz}
a^r \equiv 2 b^s \!\!\pmod{n}.
\end{equation}

We can do it by
computing $a^r \bmod{n}$ for all $r$, sorting the list of results,
and then computing $2b^s \bmod{n}$ for each $s$ one at a time, and
checking if the result appears in the sorted list.

This decreases
the time complexity of the testings phase to $O(D\log{D})$ (with the
space complexity $O(D)$).

\section{Implementation issues and improvements} \label{impl}

The theoretic base for the extension of Wiener's attack is Worley's theorem
on Diophantine approximations of the form (\ref{eq:zzz}).
We have already mentioned a result from \cite{DuIbr} which shows
that Worley's result is in some sense the best possible. However, some improvements are
possible if we consider unsymmetrical variants of Worley's result
(with different bounds on $r$ and $s$). Roughly speaking, in
solutions of (\ref{eq:zzz}) in form (\ref{eq:zzzz}), if $r<s$ then we
may take $rs<c$ instead of $rs<2c$. Due to such unsymmetrical
results, a space-time tradeoff might be possible. The following
table shows the chance of success of our attack for various
(symmetrical and unsymmetrical) bounds on $r$ and $s$. We can see
that, with the same bound for $rs$, the better results are obtained for
smaller bounds on $r$ and larger bounds on $s$. In the implementations, this fact can be used
to decrease the memory requirements (up to factor $16$).

\begin{center}
\begin{tabular}{|@{\,}c@{\,}|@{\,}c@{\,}|@{\,}c@{\,}|}
\hline\rule{0pt}{24pt} bound for $r$ & bound for $s$ &
chance of success \\[2pt]
\hline\rule{0pt}{15pt}%
$4D$ &         $4D$ &        $98 \%$   \\
$2D$ &         $2D$ &        $89 \%$ \\
$D$ &         $D$ &        $65 \%$ \\
$D$ &         $4D$ &        $86 \%$ \\
$4D$ &         $D$ &        $74 \%$ \\
$D/2$ &         $2D$ &        $70 \%$ \\
$2D$ &         $D/2$ &        $47 \%$ \\
$D/4$ &         $4D$ &        $54 \%$ \\
$4D$ &         $D/4$ &        $28 \%$ \\[2pt]
\hline
\end{tabular}
\end{center}

In the implementation of the proposed attack, we can use hash functions instead of sorting.
Furthermore, it is not necessary to store all bits of $a^r \bmod{n}$ in the hash table.
Indeed, values of $a^r \bmod{n}$ are from the set $\{0,1,\ldots,n\}$, and the number of $r$'s
is typically much smaller than $n$. Therefore, around $2\log_2{D}$ stored bits will suffice
in order to avoid too many accidental collisions. Note that a reasonable number of collisions
is not big problem here, since each such collision can be efficiently tested by Method I.
Hash tables can be used to take into account the condition $\gcd(r,s)=1$.
This condition was easy to use in brute-force testing of all possible pairs $(r,s)$,
but the direct application of our ``meet-in-the-middle'' variant seemingly ignores it.
But if we create rows in the hash table according to divisibility properties of exponents $r$
modulo small primes,
we may take again an advantage of this condition and speed up the algorithm up to 39\%.

We have implemented several variants of the proposed attack in PARI and C++,
and they work efficiently for values of $D$ up to $2^{30}$, i.e. for $d<2^{30}n^{0.25}$.

For larger values of $D$ the memory requirements become too demanding for ordinary computers.

The following table compares this bound with the bound of $d$ in the best known
attacks on RSA with small secret exponent based on LLL-algorithm.

\begin{center}
\begin{tabular}{|@{\quad}r@{\quad}|@{\quad}c@{\quad}|@{\quad}c@{\quad}|}
\hline\rule{0pt}{24pt} $\log_2{n}$ & $\log_2(2^{30}n^{0.25})$ &
$\log_2(n^{0.292})$ \\[2pt]
\hline\rule{0pt}{15pt}
512 &         158 &        150 \\
768 &         222 &        224 \\
1024 &         286 &        299 \\
2048 &         542 &        598 \\[2pt]
\hline
\end{tabular}
\end{center}

The attack can be also slightly improved by using better approximations to $\frac{k}{d}$,
e.g. $\frac{e}{n+1-2\sqrt{n}}$ instead of $\frac{e}{n}$. Namely,
\begin{equation} \label{eq:nej2}
\left| \frac{e}{n+1-2\sqrt{n}} - \frac{k}{d} \right| < \frac{0.1221\, e}{n\sqrt{n}} \,.
\end{equation}
Comparing (\ref{eq:nej2}) with (\ref{eq:nej1}),
we see that by replacing $\frac{e}{n}$ by $\frac{e}{n+1-2\sqrt{n}}$
we can gain the factor $4$ in bounds for $r$ and $s$,
so decreasing both, time and memory requirements.

With these improvements, for
1024-bits RSA modulus $n$, the range in which our attack can be applied
becomes comparable and competitive with best known attacks based on the LLL-algorithm.

\medskip

{\bf Acknowledgements.} The author would like to thank Vinko Petri\v{c}evi\'{c} for his
help with C++ implementation of the various variants
of the attack described in this paper.
The author was supported by the Ministry of Science, Education and
Sports, Republic of Croatia, grant 037-0372781-2821.

\bigskip

{\small \noindent
Andrej Dujella \\
Department of Mathematics \\ University of Zagreb
\\ Bijeni\v{c}ka cesta 30 \\
10000 Zagreb, Croatia \\
{\em E-mail address}: {\tt duje@math.hr}}

\end{document}